\newcommand\pb{PbNi$_2$V$_2$O$_8$}
\newcommand\sr{SrNi$_2$V$_2$O$_8$}
\newcommand\PbNiMgVO%
{Pb(Ni${}_{1-x}$Mg${}_x$)${}_2$V${}_2$O$_8$}
\documentstyle[prb,aps]{revtex}
\begin{document}
\title{Magnetic excitations in coupled Haldane spin chains near the quantum critical point}
\author{A. Zheludev}
\address{Physics Department, Brookhaven National Laboratory, Upton, NY
11973-5000, USA.}
\author{T. Masuda,$^*$ I. Tsukada,$^\dag$ Y.
Uchiyama,$^\ddag$ and K. Uchinokura,$^*$}
\address{Department of Applied Physics,
The University of Tokyo, 6th Engineering Bld., 7-3-1 Bunkyo-ku,
Tokyo 113-8656, Japan.}
\author{P. B\"oni}
\address{Paul Scherrer Institut, 5232 Villigen PSI, Switzerland.}
\author{S.-H. Lee}
\address{NIST Center for Neutron Research, National Institute of Standards and
Technology, MD 20899, USA.}
\address{$^\ast$ Present address: also Department of Advanced Material
Science, the University of Tokyo.\\ $^\dag$ Present address:
Central Research Institute of Electric power Industry, 2-11-1,
Iwato kita, Komae-shi, Tokyo 201-8511, Japan.\\ $^\ddag$ Present
address: ULSI Device Development Laboratory, NEC Corporation.}

\date{\today}
\maketitle
\begin{abstract}
Two quasi-1-dimensional $S=1$ quantum antiferromagnetic materials,
\pb\ and \sr, are studied by inelastic neutron scattering on
powder samples. While magnetic interactions in the two systems are
found to be very similar, subtle differences in inter-chain
interaction strengths and magnetic anisotropy are detected. The
latter are shown to be responsible for qualitatively different
ground state properties: magnetic long-range order in \sr\ and
disordered ``spin liquid'' Haldane-gap state in \pb.
\end{abstract}

\section{Introduction}
After two decades of intensive theoretical and experimental
studies, 1-dimensional (1D) Heisenberg quantum antiferromagnets
(AF) are now rather well understood. A great deal of work has been
done on integer-spin systems that have a spin-liquid ground state
and a famous Haldane gap in the magnetic excitation spectrum.
\cite{Haldane83,Haldane83-2} The focus in quantum magnetism has
now shifted towards studies of more complex phenomena, that
include inter-chain interactions, spin-lattice coupling, and/or
spin-vacancies and substitutions. Of particular current interest
is the quantum phase transition between spin-liquid (non-magnetic)
and ordered states. This type of transition in gapped 1D systems
occurs as 3D magnetic interactions and/or magnetic anisotropy are
increased beyond certain threshold values. Their effect is to
lower the energy of  excitations at certain points in reciprocal
space, and ultimately induce to a soft-mode transition to a
N\'{e}el-like ordered structure. An example of such behavior is
found in the extensively studied CsNiCl$_3$ compound.
\cite{BuyersCsNiCl3,Morra88,KakuraiCsNiCl3,Zaliznyak94} The
corresponding phase diagram has been worked out by several
authors, including Sakai and Takahashi\cite{Sakai90}
(Fig.~\ref{phase}).

The most direct way to observe such a transition experimentally is
by looking at a series of isostructural compounds with slightly
different inter-chain coupling constants or anisotropy terms. The
problem is that most known quasi-1D $S=1$ AF materials are
otherwise deep inside the spin-liquid area of the phase diagram
(good 1-D systems), or obviously in the 3D N\'{e}el-like or
XY-like ordered phases (Fig.~\ref{phase}). CsNiCl$_3$ and related
compounds\cite{BuyersCsNiCl3,KakuraiCsNiCl3,Tun91} are perhaps the
only systems close to the phase boundary that have been
extensively studied to date. Unfortunately, these compounds order
in 3 dimensions at low temperature, and have no isostructural
counterpart with a spin-liquid ground state. Only about a year ago
the first quasi-1-D integer-spin AF that is still in the
spin-liquid state, but is on the verge of a 3D ordering
instability, was characterized.\cite{Uchiyama99} This Haldane-gap
material, \pb\, is so close to the phase boundary that LRO, absent
in the pure compound, can be induced by spin-vacancy
substitution.\cite{Uchiyama99} Moreover, an isostructural undoped
system, namely  \sr\, {\it does} order in 3-dimensions at low
temperatures. For the first time two very similar stoicheometric
quasi-1D materials with such vastly different ground state
properties can be compared in experimental studies.

The magnetic properties of both \pb\ and \sr\ are due to spin
$S=1$ octahedrally-coordinated Ni$^{2+}$ ions, while the V$^{5+}$
sites are presumed to be non-magnetic. The crystal structure,
visualized in Fig.~\ref{struct}a, is tetragonal, space group $I41
cd$, with lattice constants $a = 12.249(3)$~\AA, $c =
8.354(2)$~\AA\ for \pb,\cite{Uchiyamaunpublished} and $a =
12.1617$~\AA, $c = 8.1617$~\AA\ for \sr,\cite{Wickman86}
respectively. The magnetic Ni-sites are arranged in peculiar
spiral-shaped chains that run along the unique crystal axis, as
shown in Fig.~\ref{struct}b. Even though all nearest neighbor
Ni-Ni bonds are crystallographically equivalent, the spin-spirals
have a step-4 periodicity. The dominant magnetic interaction is
antiferromagnetic, between nearest-neighbor Ni$^{2+}$ spins within
each chain. The corresponding exchange constant, $J\approx
8.2$~meV in both systems, was deduced from the high-temperature
part of the experimental $\chi(T)$ curves.\cite{Uchiyama99} The
ground state of \pb\ is a Haldane singlet, and the excitation
spectrum has an energy gap, as unambiguously shown in
low-temperature $\chi(T)$ and $C(T)$ measurements. The energy
gaps, 1.2~meV and 2.2~meV, for excitations polarized along, and
perpendicular to the chain-axis, respectively, were accurately
determined in high-field magnetization studies. The observed
anisotropy of the spin gap is attributed to single-ion easy-axis
magnetic anisotropy on the Ni-sites $D\approx -0.23$~meV.

Unlike \pb, \sr\ orders magnetically in three dimensions at
$T_{\rm N}=7$~K. The magnetic structure has not been determined to
date, but, according to bulk measurements, is of a
weak-ferromagnetic type, with a dominant antiferromagnetic
component. The ordered staggered moment is along the unique $c$
crystallographic axis. The weak-ferromagnet distortion of this
N\'{e}el (collinear) spin arrangement is attributed to the
presence of weak Dzyaloshinskii-Moriya off-diagonal exchange
interactions in the non-centric crystal.

Preliminary inelastic neutron scattering studies\cite{Uchiyama99}
provided an estimate for the inter-chain interaction strength.
Only a limited amount of neutron data are available for \pb, and,
to date, none for \sr. The present paper deals with more extensive
comparative inelastic neutron scattering studies of both
materials. Our results reveal the subtle differences between the
two systems, responsible for their vastly distinct ground state
properties.

\section{Magnetic interactions}
Prior to reporting our new experimental findings, we shall briefly
discuss the magnetic interactions that may play an important role
in the physics of \pb\ and \sr, and construct a model spin
Hamiltonian for these systems. In doing so, we shall introduce the
notation used throughout the rest of the paper.

As mentioned in the introduction, all previous studies point to
that the dominant magnetic interaction is the antiferromagnetic
coupling $J$ between nearest-neighbor spins (2.78 \AA) within each
spiral-shaped chain. It is also clear that taking the expectedly
weaker inter-chain interactions into account is crucial to
understanding the static and dynamic properties. In our previous
work (Ref.~\onlinecite{Uchiyama99}) we assumed that the dominant
inter-chain coupling is between pairs of Ni$^{2+}$ ions from
adjacent chains, such that both interacting spins have the same
$c$-axis fractional cell coordinate, and that each Ni-site is
coupled to 4 other sites. The resulting coupling topology is
schematized in Fig.~\ref{topo}a. Considering only these
inter-chain bonds, however, can not result in a very realistic
model. First of all, they do not correspond to the shortest
inter-chain Ni-Ni distance. More importantly, in the crystal
structure of \pb, there seem to be no obvious superexchange
pathways corresponding to these links. If one pays close attention
to crystal symmetry, bond length, and possible superexchange
routes involving the VO$_4$ tetrahedra, one arrives at a more
complex interaction geometry, visualized in Fig.~\ref{coupling}.
Here one assumes interactions along the shortest inter-chain Ni-Ni
bond (5.0 \AA). Coupled Ni$^{2+}$ ions are offset relative to each
other by $c/4$ along the chain axis, and are bridged by a VO$_4$
tetrahedron. All such bonds are
 crystallographically equivalent. Each Ni-site is linked with
only two adjacent chains.  The overall coupling topology for this
model is as shown in Fig.~\ref{topo}b. We shall denote the
corresponding exchange constant as $J_1$.

Magnetic anisotropy is clearly manifest in bulk susceptibility and
magnetization measurements, and must be explicitly included in the
spin Hamiltonian. Because of the strong dispersion along the
chain-axis, all the low-energy response of each chain is
concentrated at wave vectors close to the 1-D AF zone-center. In
this long-wavelength limit, 2-ion anisotropy of in-chain
interactions and single-ion anisotropy associated with individual
spins can not be distinguished. In our model we shall therefore
include only the latter, and write the corresponding term in the
Hamiltonian as:
\begin{equation}
 \hat{H}_{\rm single-ion}=D\sum_{i,k}(S_{i,k}^{z})^2.
\end{equation}
The choice of the anisotropy axis along the chain direction is
based on previous bulk magnetic studies. Here $D$ is the
anisotropy constant ($D<0$ = easy-axis) and $\bbox{S}_{i,k}$ is
the spin operator for site $i$ in chain $k$. At the same time we
shall assume in-chain exchange interactions to be isotropic and
write this Heisenberg term as:
\begin{equation}
\hat{H}_{\rm in-chain}=J\sum_{i,k}\bbox{S}_{i,k}\bbox{S}_{i+1,k}.
\end{equation}
The magnitude of dispersion perpendicular to the chain axis is
expected to be rather small, so the entire range of wave vector
transfers in the $(a,b)$ plabe will influence the low-energy
properties. For this reason, two-ion anisotropy of $J_1$, unlike
that of $J$, is a relevant parameter and should be considered. The
inter-chain coupling term in the Hamiltonin will thus have the
form:
\begin{equation}
 \hat{H}_{\rm inter-chain}  =
 \sum_{i,i',k,k'}\left[J_{1,\|}S_{i,k}^{z}S_{i',k'}^{z}+J_{1,\bot}\left\{S_{i,k}^{x}S_{i',k'}^{x}+S_{i,k}^{y}S_{i',k'}^{y}\right\}\right].
\end{equation}
Here the sum is taken over pairs of next-nearest-neighbor spins.
The three terms discussed above constitute the spin Hamiltonian
that we propose for \pb\ and \sr. Of course, other magnetic
interactions can be active in the system as well. As will be
discussed below, however, the described model can reproduce the
experimental data with a minimal number of parameters.

\section{Experimental}
Inelastic neutron scattering studies of \pb\ and \sr\ powder
samples (about 10 g each) were carried out in two series of
experiments. The conventional 3-axis technique was used to measure
certain characteristic constant-$E$ and constant-$Q$ scans. These
measurements were performed at the TASP and Druchal spectrometers
at the continuous spallation source SINQ at Paul Scherrer
Institute. Neutrons of a fixed final energy of 5~meV or 8~meV were
used with pyrolytic graphite (PG) monochromator and analyzer, and
a typical $\text{(open)}-80'-80'-\text{(open)}'$ collimation
setup. For some scans a horizontally focusing analyzer was used to
increase the useful scattered intensity, in which case no
collimators were used after the sample. In most of the
measurements a flat analyzer was employed instead. In the
5~meV-final configuration a Be filter was inserted after the
sample to suppress higher-order beam contamination. The spectrum
of neutrons emerging from the guide is such that no higher-order
filter was required for the 8~meV-final measurements.

In the second series of experiments we took advantage of the
area-sensitive detector (ASD) 3-axis setup available for the NG-5
``SPINS'' spectrometer installed at the National Institute of
Standards and Technology. In special cases when a large domain of
$E-Q$ space needs to be surveyed in a spherically symmetric sample
(such as powder) this technique can provide an amazing, almost
10-fold increase of data collection rate, as compared to the
standard 3-axis geometry, without any penalty in resolution. In
our experiments on \pb\ and \sr\ we utilized a PG monochromator
and a $\text{(open)}-80'-80'-80'{\rm (radial)}$ array of
collimators. The measurements were done in the fixed-final-energy
mode, with the central blade of the composite analyzer tuned to
$E_{\rm f}=3.125$~meV. In the experiment scattering events with
final neutron energies in the range $E_{\rm f}\pm 0.4$~meV are
registered simultaneously. To suppress higher-order beam
contamination we used a Be-O filter after the sample.  The data
were taken for momentum and energy transfers of up to
2.5~\AA$^{-1}$ and 7~meV, respectively.

In both series of experiments the sample environment was a
standard ``ILL-orange'' cryostat, and the temperature range
$1.5-30$~K was covered. The background was measured by repeating
some scans with the analyzer moved away from its elastic position
by 10$^\circ$. In the SPINS experiment the background due to air
scattering was separately measured with the sample removed from
the spectrometer.

\section{Experimental results}
\subsection{Results obtained at low temperatures}
The inelastic neutron scattering intensities measured in \pb\ and
\sr\ at $T=1.2$~K  with the ASD setup are shown in the false-color
plot in Fig.~\ref{ASDdata} (background subtracted). The resolution
of the area-sensitive detector is greater than the actual energy
and wave vector resolution of the spectrometer (0.11~meV and
0.023~\AA$^{-1}$ at $\hbar \omega=0$). The customary procedure is
to re-bin the mesh data to a coarser grid. Instead, for
visualization purposes, the data in Fig.~\ref{ASDdata} were
smeared using a fixed-resolution 0.5~meV$\times$0.06~\AA$^{-1}$
FWHM Gaussian filter. As reported previously,\cite{Uchiyama99}
constant-energy scans measured in \pb\ at 4~meV energy transfer
have a peculiar and very characteristic shape. Such scans were
collected at $T=2$~K for both materials, using the
8~meV-final/flat analyzer setup SINQ. Theses scans are shown in
Fig.~\ref{conste}. In addition, representative constant-$Q$ scans
were also measured in this configuration, and are shown in
Figs.~\ref{constqpb} and~\ref{constqsr}.

Despite the different ground states, the main features of the
powder-averaged dynamic cross section for the two materials are
quite similar. This, however, is not particularly surprising. The
dynamic structure factor $S(\bbox{Q},\omega)$ is severely smeared
out by the spherical averaging that occurs in a powder sample. The
main features are defined by the form factor of each spiral-shaped
chain, and by the steep dispersion along the chain axis, expected
to be almost identical in the two systems. The significant
difference between the two materials is expected to be in the
magnitude of the inter-chain interactions, and possibly single-ion
anisotropy. These effects are much more difficult to observe, as
they only influence the weak spin wave dispersion in the $(a,b)$
crystallographic plane. Our data do in fact contain relevant
information on the transverse dispersion of spin excitations,
which can be extracted from a quantitative analysis, as described
in the following section.

\subsection{Data analysis}
\subsubsection{Model cross section}
The general problem in interpreting inelastic neutron scattering
data from  powder samples is an effective loss of information upon
spherical averaging. Indeed, the quantity of interest is the
dynamic structure factor $S(\bbox{Q},\omega)$ that for each
channel of spin polarization is a scalar function in 4-dimensional
$E-\bbox{Q}$ space. In a powder sample one measures the spherical
average of this function, a scalar function defined in
2-dimensional space:
\begin{equation}
S_{\text{powder}}(Q,\omega)= \frac{1}{4\pi} \int d\phi \sin\theta
d\theta S([Q\sin\theta \cos\phi,
Q\sin\theta\sin\phi,Q\cos\theta],\omega)
\end{equation}
The transformation $S(\bbox{Q},\omega) \rightarrow
S_{\text{powder}}(Q,\omega)$ is not reversible. The only way the
full correlation function  can be extracted from the experiment is
by assuming some parameterized model for $S(\bbox{q},\omega)$ and
fitting it to the measured $S_{\text{powder}}(Q,\omega)$. For \pb\
and \sr\ we used the following analytical form for
$S(\bbox{q},\omega)$, as derived below in the theory section:
\begin{eqnarray}
 S(\bbox{Q})&  = & P_{\|}(\bbox{Q})S_{\|}(\bbox{Q})+ P_{\bot}(\bbox{Q})S_{\bot}(\bbox{Q}),\label{sfit}\\
 2S_{\|}(\bbox{Q})& = &  2\cos^2\psi_1\cos^2\psi_2S'_{\|}(h,k,l)\nonumber\\
 & + &  \cos^2\psi_1\sin^2\psi_2
 \left[ S'_{\|}\left(h+1,k,l+1\right)\right.
 \left.+S'_{\|}\left(h+1,k,l+3\right)\right]\nonumber\\
 & + &  \sin^2\psi_1\cos^2\psi_2
 \left[ S'_{\|}\left(h,k+1,l+1\right)\right.
 \left.+S'_{\|}\left(h,k+1,l+3\right)\right]\nonumber\\
 & + & 2\sin^2\psi_1\sin^2\psi_2
 S'\left(h+1,k+1,l+2\right),\label{traneq1}\\
 2S_{\bot}(\bbox{Q}) & = & 2\cos^2\psi_1\cos^2\psi_2S'_{\bot}(h,k,l)\nonumber\\
 & + & \cos^2\psi_1\sin^2\psi_2
 \left[ S'_{\bot}\left(h+1,k,l+1\right)\right.
 \left.+S'_{\bot}\left(h+1,k,l+3\right)\right]\nonumber\\
 & + &  \sin^2\psi_1\cos^2\psi_2
 \left[ S'_{\bot}\left(h,k+1,l+1\right)\right.
 \left.+S'_{\bot}\left(h,k+1,l+3\right)\right]\nonumber\\
 & + & 2\sin^2\psi_1\sin^2\psi_2
 S'\left(h+1,k+1,l+2\right),\\
 \psi_1 &=&\frac{\pi d}{a}h,\\
 \psi_2 &=&\frac{\pi d}{a}k.\label{traneq2}
\end{eqnarray}
In these formulas the argument $\omega$ has been dropped. The
phases $\psi_1$ and $\psi_2$ represent the 3-D structure factor of
the spiral-shaped spin chains, and $d=0.08a$ is the offset of each
Ni$^{2+}$-ion relative to the central axis of the corresponding
spiral chain, along the $a$ or $b$ axis. The polarization factors
for longitudinal (polarization along the $c$-axis) and transverse
(polarization in the $(a,b)$ plane) spin excitations are defined
as:
\begin{eqnarray}
P_{\|}(\bbox{Q})=\sin^2(\widehat{\bbox{Q},\bbox{z}}),\\
P_{\bot}(\bbox{Q})=1+\cos^2(\widehat{\bbox{Q},\bbox{z}}).
\end{eqnarray}
The dynamic structure factors $S'_{\|}$ and $S'_{\bot}$ for
straight (as opposed to spiral-shaped) Haldane chains are written
in the Single-Mode Approximation (SMA)
\cite{Arovas88,Muller81,Ma92}:
\begin{eqnarray}
 S'_{\|}(\bbox{Q},\omega)=\frac{Zv}{\hbar
 \omega_{\|}(\bbox{Q})}\delta(\hbar \omega- \hbar
 \omega_{\|}(\bbox{Q})),\label{badd1}\\
 S'_{\bot}(\bbox{Q},\omega)=\frac{Zv}{\hbar
 \omega_{\bot}(\bbox{Q})}\delta(\hbar \omega- \hbar
 \omega_{\bot}(\bbox{Q})).\label{badd2}
\end{eqnarray}
Here $v=2.48J$ is the spin wave velocity, and $Z=1.26$
(Ref.~\onlinecite{Sorensen94}). Finally, the dispersion relation
for weakly coupled chains are given by:
\begin{eqnarray}
 \left[\hbar \omega_{\|}(\bbox{Q})\right]^2 & = &
 \Delta_{\|}^2+v^2\sin^2(\pi l/2)
 -\frac{1}{2}ZvJ_{1,\|}\cos(\pi l/2)\left[ \cos(\pi h) +\cos(\pi k)\right](1-\cos \pi l/2),
  \label{disp1}\\
 \left[\hbar \omega_{\bot}(\bbox{Q})\right]^2  & = &
 \Delta_{\bot}^2+v^2\sin^2(\pi l/2)
 -\frac{1}{2}ZvJ_{1,\bot}\cos(\pi l/2)\left[ \cos(\pi h) +\cos(\pi k)\right](1-\cos \pi
 l/2),\label{disp2}
\end{eqnarray}
where $\Delta_{\|}$ and $\Delta_{\bot}$ are the longitudinal and
transverse {\it intrinsic Haldane} gaps for non-interacting
chains, respectively. The splitting of the triplet in our model is
caused by single-ion anisotropy:\cite{Golinelli92}
\begin{eqnarray}
 \Delta_{\bot}=\langle \Delta \rangle-0.57 D,\label{aniso1}\\
 \Delta_{\|}=\langle \Delta \rangle+1.41 D\label{aniso2}.
\end{eqnarray}
An important parameter is the {\it mean} intrinsic Haldane gap
$\langle \Delta \rangle \equiv(\Delta_{\|}+2\Delta_{\bot})/3$. As
can be seen from Eqs.~\ref{aniso1} and \ref{aniso2}, it is, to the
first order, defined by $J$ alone: $\langle \Delta \rangle\approx
0.41 J$.\cite{Golinelli92} It is also useful to define the actual
gaps (excitation energies at the 3-D AF zone-center):
\begin{eqnarray}
 E_{\text{min},\bot}^2  =
 \Delta_{\bot}^2-2Zv|J_{1,\bot}|,\label{Emin1} \\
 E_{\text{min},\|}^2  =
 \Delta_{\|}^2-2Zv|J_{1,\|}|.\label{Emin2}
\end{eqnarray}

The spherical average of Eq.~\ref{sfit} was calculated numerically
using a Monte-Carlo algorithm that also eliminates the
$\delta$-functions in Eqs.~\ref{badd1} and \ref{badd2}. The
parameters were then refined by a standard least-squares routine
to best-fit the data. To accelerate the fitting process the ASD
data were binned to a rectangular 0.046~\AA$^{-1}\times$ 0.1~meV
resolution. An additional benefit of this binning is that it
allowed us not to worry about resolution effects. The lower
0.5~meV energy transfer range, that contains the
elastic-incoherent and possibly phonon scattering, was excluded
from the fits.

\subsubsection{Analysis of \pb\ data.}
For The Pb-compound the adjustable parameters were the two
intrinsic gap energies $\Delta_{\|}$ and $\Delta_{\bot}$, and the
doublet 3D gap $ E_{\text{min},\bot}$. The singlet 3-D gap was
fixed to $E_{\text{min},\|}=1.2$~meV, as determined in high-field
bulk measurements.\cite{Uchiyama99} The only additional parameter
was an overall scale factor. The least-squares refinement yields:
$\Delta_{\bot}=4.0 \pm 0.25$~meV, $\Delta_{\|}=3.1 \pm 0.3$~meV
and $E^{\rm (min)}_{\bot}=2.4 \pm 0.2$~meV  ($J_1<0$), with
$\chi^2=2.3$. The resulting fit is shown in Fig.~\ref{simpb}.
Substituting the  obtained values values into the expression for
$S({\bbox{Q},\omega})$, performing a powder average and
convoluting the result with the spectrometer resolution function,
reproduces the measured 3-axis const-$E$ and const-$Q$ scans
rather well, as shown in solid lines in
Figs.~\ref{conste}b,\ref{constqpb}. The relatively large $\chi^2$
of the global fit to the ASD data is to be attributed to
systematic error, primarily due to ``spurious'' scattering. In
particular, the data are partially contaminated by spurions of
type ``$k_i \rightarrow k_i$'' and ``$k_f \rightarrow k_f$''
originating from the stronger Bragg powder lines from the sample.
Areas that are affected by these spurious processes are shown as
shaded ``streaks'' in Fig.~\ref{ASDdata}. Another prominent streak
in the data is to the left from, and parallel to, the two shaded
areas. The origin of this feature is not known. It is however
resolution-limited and temperature-independent, and is thus almost
certainly spurious and of non-magnetic origin. Considering that
this type of systematic error is unavoidable in powder
experiments, the obtained model fit to the data is quite
acceptable.

The refined value for $E_{\text{min},\bot}$ is in very good
agreement with the result of high-field
measurements.\cite{Uchiyama99} The interchain coupling constants
are obtained from Eqs.~\ref{Emin1},\ref{Emin2}:
$J_{1,\bot}=-0.18$~meV and $J_{1,\|}=-0.14$~meV. Note that $|J_1|$
is larger by a factor of 1.5--2, compared to our previous estimate
($J_{\bot}=0.096\pm 0.003$~meV) in Ref.~\onlinecite{Uchiyama99}.
This discrepancy should be partly attributed to a difference in
the definition of $J_1$. In our previous model each spin was
coupled to 4 spins in adjacent chains (coordination number 4). In
the present model the inter-chain coordination number is 2. This
automatically translates into a factor of 2 for $|J_1|$. The
negative sign of $J_1$ indicates that this coupling is actually
ferromagnetic. Note, however, that in our model $J_1$ couples
spins offset by $c/4$ along the chain axis, so that the effective
mean field coupling between interacting chains is still
antiferromagnetic, as in our previous model. The 3-D magnetic
zone-center, where transverse dispersion is a minimum, is at
$(1,1,2)$. From the refined $\Delta_{\bot}$ and $\Delta_{\|}$ we
can also get the in-chain coupling constant: $J\approx9.0$~meV.
This value is in a better agreement with the high-temperature
susceptibility estimate $J\approx 8.2$~meV in
Ref.~\onlinecite{Uchiyama99}, than our preliminary neutron result
$J=9.5$~meV, from the same reference. The anisotropy constant $D$
can be estimated from $\Delta_{\bot}$ and $\Delta_{\|}$ using
Eqs.~\ref{aniso1},\ref{aniso2}. For \pb\ we get $D=-0.45$~meV.

To convince the reader that the $J_1$ is in fact a relevant
parameter, we have performed simulations of the dynamic structure
factors using the same values $\Delta_{\bot}=4.0 \pm 0.25$~meV,
$\Delta_{\|}=3.1 \pm 0.3$~meV, as determined in the analysis of
the ASD data, but with the sign of $J_1$ reversed:
$J_{1,\bot}=0.18$~meV and $J_{1,\|}=0.14$~meV. The resulting
simulated powder cross section is visualized in
Fig.~\ref{simpbbad}a, and is clearly very different from the
inelastic intensity measured experimentally. Similarly,
Fig.~\ref{simpbbad}b shows a simulation with
$J_{1,\bot}=J_{1,\|}=0$~meV. $J_1$ has a particular impact on the
shape of the constant-energy powder scans. The dashed lines in
Fig.~\ref{conste} are simulations for uncoupled chains. The
well-defined intensity maximum seen in the data at
$|Q|\approx1.2$~\AA$^{-1}$, is replaced with a broad monotonous
feature for uncoupled chains. In contrast, the interacting chain
model reproduces the peak rather well.

\subsubsection{Analysis of \sr\ data} Similar data analysis was performed for \sr. $E^{\rm
(min)}_{\|}$ in this case was fixed at zero value (see discussion
in the theory section below). The best fit to the ASD data is
obtained with $\Delta_{\bot}=3.9 \pm 0.3$~meV, $\Delta_{\|}=2.8
\pm 0.4$~meV and $E^{\rm (min)}_{\bot}=2.35 \pm 0.3$~meV with
$\chi^2=2.0$, and is shown in Fig.~\ref{simpb}b. These values
correspond to interchain coupling constants $J_{1,\bot}=-0.18$~meV
and $J_{1,\|}=-0.15$~meV, respectively, almost identical to the
corresponding \pb\ values. An important consistency check is that
using the gap values to estimate $J$ gives almost the same value
as for \pb, as expected: $J=8.6$~meV. The easy-axis anisotropy
constant in \sr\ is larger than in \pb: $D=-0.56$~meV. Simulations
for constant-$E$ and constant-$Q$ scans measured for \sr\ in the
standard 3-axis mode are shown in solid lines in
Figs.~\ref{conste}a,\ref{constqsr}.

\subsubsection{Observation of actual spin gaps at the 3D magnetic
zone-center}

From the point of view of spin dynamics, the main distinction
between the singlet-ground-state \pb\ and the magnetically ordered
\sr\ is the presence of a spin gap in the former system, and a
gapless excitation spectrum in the latter. The gap is directly
accessible experimentally at the 3-D AF zone-center, where the
dispersion of magnetic excitations is a global minimum. As
mentioned above, for both vanadates the 3D zone-center is at
$\bbox{Q}^{(0)}=(1,1,2)$, which corresponds to a momentum transfer
$|\bbox{Q}^{(0)}|\approx1.67$~\AA$^{-1}$. Constant-$Q$ scans
extracted from our ASD data for this momentum transfer, as well as
a standard  $E_f=5$~meV constant-$Q$ scan at this wave vector, are
shown in Fig.~\ref{gapnogap}. To better understand these data, we
note that near the 3D zone-center the single-mode contribution to
the dynamic structure factor, independently of the details of the
spin Hamiltonian, should be of the following form:
\begin{eqnarray}
S(\bbox{Q},\omega) &\propto&
\frac{1}{\omega_{\bbox{Q}}}\delta(\omega-\omega_{\bbox{Q}}),
\label{emp}\\
 (\hbar
\omega_{\bbox{Q}})^2&=&\Delta^2+v_{\|}^2(Q_{\|}-Q^{(0)}_{\|})^2+v_{\bot}^2(Q_{\bot}-Q^{(0)}_{\bot})^2.\nonumber
\end{eqnarray}
In this formula $\Delta$ is the gap energy, and $v_{\|}$ and
$v_{\bot}$ are spin wave velocities along and perpendicular to the
chain axis $c$, respectively. The data shown in
Fig.~\ref{gapnogap}c were collected with a horizontally-focusing
analyzer. At energy transfers below 1~meV in this mode we are
picking up a great deal of diffuse and phonon scattering (
$(1,1,2)$ is an allowed nuclear Bragg peak !). Above this
contaminated region though,  in a powder sample we effectively
observe a $\bbox{Q}$-integral of the cross-section around the 3D
zone-center. The same applies to the scans in
Figs.~\ref{gapnogap}a and b, where integration was performed in
the range 1.6--1.7~\AA$^{-1}$. For the $Q$-integrated intensity
Eq.~\ref{emp} gives:
\begin{equation}
S(\omega)\propto \sqrt{\omega^2-\Delta^2}.
\end{equation}
The linear increase of intensity seen for \sr\  in
Fig.~\ref{gapnogap} should thus indeed be interpreted as due to a
gapless spin wave ($\Delta=0$). In contrast, for \pb, the
threshold behavior is a clear sign of an energy gap. These effects
are also seen in constant-$Q$ scans extracted from the ASD data by
binning the pixels in a 0.1~\AA$^{-1}$ $Q$-range.

\subsection{\sr: Temperature dependence}
To better understand the mechanism of long-range ordering of \sr\,
we studied the temperature dependence of inelastic scattering at
the momentum transfer $|\bbox{Q}^{(0)}|=1.67$~\AA$^{-1}$ in this
compound. Typical data are shown in Fig.~\ref{vstdata}. The solid
lines were obtained by fitting our model cross section to the data
at each temperature. The lower 0.75~meV energy transfer range was
excluded from this analysis, as the difficult-to-estimate phonon
contribution in this range is expected to increase dramatically
with increasing $T$. The set of independent parameters was
slightly modified. The anisotropy splitting of the triplet $D$, as
well as both inter-chain coupling constants $J_{1,\bot}$ and
$J_{1,\|}$ were fixed at the values determined at $T=1.5$~K. The
mean gap $\langle \Delta \rangle$ and an intensity prefactor were
refined to best-fit the scans at each temperature. The obtained
temperature dependences are shown  in Fig.~\ref{vstresult}. As
previously observed in other Haldane-gap systems, $\langle
\Delta\rangle$ increases with increasing $T$. Even though this
change is rather small, according to Eqs.~\ref{Emin1},\ref{Emin2},
it corresponds to an appreciable variation of the gap in the
longitudinal mode (Fig.~\ref{vstresult2}). Upon cooling, the
longitudinal gap approaches zero at $T=T_{\rm N}$, which results
in a soft-mode transition to a magnetically ordered state. This
type of behavior is very similar to that found in CsNiCl$_3$.
\cite{BuyersCsNiCl3,Morra88,KakuraiCsNiCl3}

\section{Theory and discussion}
\subsection{Derivation of the model cross section}
To calculate the effect of  weak inter-chain interactions on
dynamic spin correlations in \pb\ and \sr\ we shall use the Random
Phase Approximation (RPA).\cite{Scalapino75} This approach for
directly coupled Haldane spin chains has been successfully
applied, for example, to CsNiCl$_3$ (Ref.~\onlinecite{Morra88}).
More recently it was shown to also work well for Haldane chains
coupled via classical spins.\cite{Zheludev99NBANO} The only
additional difficulties in the present case arise from the
complicated 3-D arrangement of magnetic ions in the \pb\
structure, and from the rather non-trivial geometry of interchain
coupling. To somewhat simplify the task we shall break it up into
two distinct problems. First, we shall worry only about the {\it
topology} of magnetic interactions and consider an equivalent
Bravais lattice of spins, assuming straight spin chains and a
system of inter-chain bonds shown in Fig.~\ref{topo}b. Making use
of the general RPA equations for coupled spin chains, summarized
in Appendix I, we shall write down the RPA susceptibility and
dynamic structure factor $S'(\bbox{Q},\omega)$ for this model. In
a separate step we shall adapt the result to the more complex
structure of \pb, using the formulas of Appendix II.

\subsubsection{Structure factor for an equivalent Bravais lattice}
The first step in the RPA calculation is to write down the bare
(non-interacting) susceptibility for an isolated Haldane spin
chain, for which the single-mode approximation (SMA) is known to
work rather well:\cite{Arovas88,Muller81,Ma92}
\begin{equation}
 \chi_0(\bbox{Q},\omega)=\frac{1-\cos (Q_zc/4)}{2}\frac{Zv}{\Delta^2+v^2\sin^2(Q_zc/4)-(\hbar
 \omega+i\epsilon)^2}\label{bare}
\end{equation}
This expression should be used separately for each channel of spin
polarization, with appropriate values of gap energy $\Delta$ for
each particular mode.

Expression \ref{bare} is to be substituted in the general RPA
equations \ref{self1}. Even for the ``straightened out'' spin
chains in Fig.~\ref{topo}b there are still 16 spins per unit cell,
so we end up with 16 couple equations and 16 modes for each
polarization! For the low-energy part of the excitation spectrum
it is however quite appropriate to use the approximation of
Eqs.~\ref{self2}. This reduces the problem to only 4
self-consistent RPA equations (there are 4 spin chains in each
crystallographic unit cell). Moreover, at this level of
approximation the set of 4 RPA equations is degenerate, and we end
up with a single RPA equation for each polarization:
\begin{equation}
 \chi^{-1}_{\rm RPA}(\bbox{Q})  =  \chi_0^{-1}(\bbox{Q})
  \left[\chi_{0}(\bbox{Q}) {\cal J}(\bbox{Q})+1\right],
\end{equation}
where
\begin{equation}
 {\cal J}(\bbox{Q})=J_1\cos(Q_zc/4)\left[ \cos(Q_xa/2) +\cos(Q_ya/2)\right].
\end{equation}
This equation is easily solved analytically. By taking the
imaginary part of the thus obtained $\chi_{\rm RPA}(\bbox{Q})$ we
arrive at Eqs.~\ref{badd1}--\ref{disp2}.

\subsubsection{Actual structure} The transformation from the
Bravais spin lattice (straight chains) to the spiral-chain
structure of \pb\ and \sr\ is  schematized in
Fig.~\ref{transformfig}. This figure also serves as an
illustration of the notation used in Appendix I. Applying
Eq.~\ref{transform} is straightforward, and we will spare the
reader the tedious calculations. The final relation between
$S(\bbox{q},\omega)$ and $S'(\bbox{q},\omega)$ is as in
Eqs.~\ref{traneq1},\ref{traneq2}.

\subsection{Temperature effects}
\subsubsection{Disordered phase}
The derivation above can be repeated almost verbatim for the case
$T>0$. Temperature enters  RPA calculation indirectly, through an
intrinsic temperature dependence of bare susceptibilities of
individual chains (Eq.~\ref{bare}). Two effects at $T>0$ need be
considered: i) the increase of the Haldane gap, compared to its
$T=0$ value, and ii) damping of Haldane excitations. Both
phenomena have been observed in a number of model quasi-1-D
systems (see, for example,
Refs.~\onlinecite{Ma95,Zheludev96-NINAZ,Sakaguchi96}).
Unfortunately, it is almost impossible to extract meaningful
information regarding excitation lifetimes from powder inelastic
data. As most of the observed inelastic signal originates from
excitations with energies greater than 2~meV, the temperature
dependence of the intrinsic Haldane gap is much easier to observe.
For this reason we analyzed the temperature dependence of the
inelastic signal measured in \sr\ using the same single-mode
approximation (Eq.~\ref{bare}), as at base temperature (see fits
in Fig.~\ref{vstdata}), to obtain the $T$-dependences of intrinsic
Haldane gap energies (Fig.~\ref{vstresult}). From the mapping of
the Heisenberg Hamiltonian on the quantum non-linear
$\sigma$-model (NLSM) one expects a rather steep increase of the
gap energy with increasing temperature:\cite{Golinelli94}
\begin{equation}
\Delta(T)\approx\Delta(0)+\sqrt{2\pi}\sqrt{T\Delta(0)}\exp(-\frac{\Delta(0)}{k_{\rm
B} T})\label{NLSM}.
\end{equation}
The actual increase of the gap energy observed experimentally
NENP,\cite{Ma95} NINAZ,\cite{Zheludev96-NINAZ} and
Y$_{2}$BaNiO$_5$,\cite{Sakaguchi96} was found to be consistently
smaller that this NLSM prediction. The same discrepancy, namely a
rather slow increase of the gap energy with temperature, is seen
in \sr\ as well. Equation~\ref{NLSM} is plotted in a dashed line
in Fig.~\ref{vstresult}a. The solid line is an empirical fit in
which the prefactor $\sqrt{2\pi}$ in Eq.~\ref{NLSM} was replaced
by an adjustable parameter $A$. For \sr\ we get $A=0.67(1)\ll
\sqrt{2\pi}$. A self-consistency check for our SMA model is that
the refined intensity prefactor is almost $T-independent$, as
shown in Fig.~\ref{vstresult}b. In other words, the decrease of
actual excitation intensity is entirely due to the increase of gap
energy, and the intensity scales as $1/\omega$.

In \sr\ the ordering temperature $T_{\rm N}=7$~K is significantly
smaller than the intrtinsic Haldane gap energy
$\Delta\approx4$~meV. As both the gap energy and excitation width
are expected to increase exponentially with $T$, such a small
$T_{N}$ suggests that inter-chain interactions in \sr\ are barely
strong enough to produce a LRO-ground state. Long-range ordering
in \sr, and the absence of such in \pb, are a very ``lucky''
coincidence that results from a fine interplay between in-chain
interaction strength, magnetic anisotropy and inter-chain
coupling.

\subsubsection{Magnetically ordered state}
Our model cross section was derived under the implicit assumption
that the system is in a non-magnetic state, which is {\it not}
applicable to \sr\ at $T<T_{\rm N}$. In this regime one expects an
{\it increase} of the gap energies, at least for the longitudinal
mode, due to the presence of a mean static staggered field
generated by the ordered staggered moment in the
system.\cite{Affleck89-2,MZ98-L} To adapt the MF-RPA calculation
to this temperature regime one has to know the ordered moment in
the system. To date, powder diffraction experiments on \sr\ failed
to detect any magnetic Bragg reflections in this compound at low
temperatures,\cite{Boniunpublished} and the magnetic order
parameter is thus expected to be very small. This is consistent
with our previous observation that the system is ``barely''
3-dimensional enough to become ordered at low temperatures.
Provided the ordered moment is small, within the accuracy of our
powder experiments it seems quite appropriate to use the same
cross section for $T<T_{\rm N}$, as for the paramagnetic phase,
and postulate $E_{{\rm min}, \|}=0$. This assumption is equivalent
to assuming $T_{\rm N}=0$.

\subsubsection{Placement on the phase diagram}
The line of quantum phase transition separating the ordered and
spin-liquid states can be derived from
Eqs.~\ref{aniso1}-\ref{Emin2}. The critical value for $J_1$
corresponds to the lower gap energy (in our case $E_{{\rm
min},\|}$) being equal to zero. The resulting phase diagram is
very similar to the direct numerical calculations by Sakai and
Takashi,\cite{Sakai90} and is shown in Fig.~\ref{phase}. Using the
results for \pb\ and \sr\ described above, we are able to place
the two new materials on the same plot. It has to be emphasized,
that for \sr\  the parameters $D$ and $J_1$ were extracted from
the experimental data using a cross-section for a disordered
system, and assuming $E_{{\rm min},\|}\equiv 0$. That \sr\ lands
{\it exactly} on the phase boundary is thus an artifact of our
data analysis procedure. In reality, \sr\ must be positioned
slightly above the phase boundary. The disordered \pb\ is not
influenced by such an artifact, and its positioning can be
considered quite reliable.

\section{Summary}
Our neutron scattering results help explain the different ground
state properties of the two very similar vanadates. The ratio of
inter-chain to in-chain coupling in \sr\ is slightly larger than
in \pb. \sr\ is also driven towards LRO by the somewhat larger
magnetic anisotropy. The quantitative data analysis enables us to
precisely position the two compounds on the Sakai-Takahashi phase
diagram, just opposite to each other relative to the
ordered-disordered phase boundary. This almost unbelievable
coincidence opens many exciting possibilities for future studies.
Experiments on aligned powders will enable more accurate
measurements of the excitation spectrum near the 3D zone-center.
High-pressure studies, as previously attempted for
NENP,\cite{Zaliznyak98} may lead to the first observation of
pressure-induced long-range ordering in a quantum-disordered
magnet.

\acknowledgements We would like to thank Y. Sasago, who has
suggested at the early stage of this study that \sr\ may be a
Haldane-gap antiferromagnet. We also thank R. Wickmann for sending
a copy of Ref.~\onlinecite{Wickman86}. Work at the University of
Tokyo was supported in part by Grant-in-Aid for COE Research
``Phase Control of Spin-Charge-Phonon Coupled Systems'' from the
Ministry of Education, Science, Sports and Culture of Japan.  Work
at Brookhaven National Laboratory was carried out under Contract
No.  DE-AC02-76CH00016, Division of Material Science, U.S.\
Department of Energy. Studies at NIST were partially supported by
the NSF under contract No. DMR-9413101.

\section*{Appendix I: RPA susceptibility for weakly coupled spin chains}
The general RPA technique of calculating the dynamic
susceptibility and structure factor for weakly interacting systems
is well established and documented (see, for example,
Ref.~\onlinecite{JensenMackintosh}). In particular, the approach
has been on many occasions applied to weakly coupled quantum spin
chains.\cite{Scalapino75,Morra88}  In the present section, without
any claims of novelty, and for reference only, we shall derive
some useful RPA results for weakly coupled spin chains assuming a
rather general geometry of inter-chain bonding. Note that the
notation used in this Appendix is totally independent from that in
the rest of the paper.

We consider a crystal structure composed of identical parallel
uniform spin chains that run along the $\bbox{a}$ axis. The
spacing between spins in each chain is $a/K$, where $K$ is
integer. There are $M$ chains in each crystallographic unit cell.
The origin of the $m$-th chain is at $\bbox{R}_n+\bbox{\rho}_{m}$.
The position of the $i$-th spin in chain $(n,m)$ can be written as
$\bbox{R}_n+\bbox{\rho}_{m}+i\bbox{a}/K$. A single
crystallographic unit cell thus contains $MK$ spins. For
convenience we shall break up the in-chain spin index $i$ into
two: $i=lK+k$, $0\le k < K$. The spin Hamiltonian for inter-chain
interactions will be written as:
\begin{equation}
  \hat{H}_{\rm
  interchain}=\sum_{n,n'}\sum_{m,m'}\sum_{l,l'}\sum_{k=0}^{K-1}\sum_{k'=0}^{K-1}
  s^{(n,m)}_{l,k}s^{(n',m')}_{k',l'}J(n,n',m,m',Kl+k,Kl'+k')
\end{equation}
We assume all exchange interactions to be simultaneously diagonal
in spin projection indexes. All the equations in this section thus
apply to a particular channel of spin polarization. In our model
the exchange constant satisfies certain translational-symmetry
relations:
\begin{equation}
J(n,n',m,m',Kl+k,Kl'+k')=J((\bbox{R}_{n'}-\bbox{R}_{n}),m,m',(l'-l),k,k').
\end{equation}

In order to write down self-consistent  RPA equations, we have to
answer the following question. Suppose we have artificially
induced a spin density $s^{(m')}(\bbox{r})$ (or, equivalently,
$s^{(m')}(\bbox{q})$)in the chain-sublattice $m'$.  What exchange
field will this spin density project on the chains in sublattice
$m$? By definition, the exchange field acting on spin
$s^{(n,m)}_{l,k}$ is given by:
\begin{equation}
  h^{(n,m)}_{l,k}=\sum_{n'}\sum_{m'=0}^{M-1}\sum_{l'}\sum_{k'=0}^{K-1}
  s^{(n',m')}_{l',k'}J((\bbox{R}_{n'}-\bbox{R}_{n}),m,m'(l'-l),k,k')
\end{equation}
We, of course, will be interested in the Fourier transform of the
exchange field acting on sublattice $m$:
\begin{equation}
 h^{(m)}(\bbox{q})  \equiv \sum_{n}\sum_l\sum_{k=0}^{K-1}h^{(n,m)}_{l,k} \exp\left[-{\rm i}\bbox{q}(\bbox{R}_n+\bbox{\rho}_m+\bbox{a}(k/K+l))\right].
\end{equation}
Having introduced the definition
\begin{equation}
J_{k,m,m'}(\bbox{q})\equiv
 \sum_{n,k',l} \exp\left[ -{\rm i}\bbox{q}(\bbox{R}_n+\bbox{\rho}_{m'}-\bbox{\rho}_{m}+\bbox{a}l+\bbox{a}(k'-k)/K)\right]
 J(\bbox{R}_n,m',m,l,k',k),
\end{equation}
it is straightforward to verify that:
\begin{equation}
 h^{(m)}(\bbox{q})=\sum_{m'=0}^{M-1}\sum_{\kappa=0}^{K-1}{\cal J}_{\kappa,m,m'}(\bbox{q}) s^{(m')}(\bbox{q}+\kappa
 \bbox{a}^{\ast}),\label{field}
\end{equation}
where
\begin{equation}
 {\cal J}_{\kappa,m,m'}(\bbox{q})\equiv \frac{1}{K}\sum_{k}J_{k,m,m'}(\bbox{q}) \exp\left[2\pi{\rm i}  \kappa k/K\right].
\end{equation}
Equation \ref{field} enables us to immediately write down the
self-consistent RPA equations:
\begin{equation}
 \chi^{(m)}(\bbox{q},\omega)=
  \left[\sum_{m'=0}^{M-1} \sum_{\kappa=0}^{K-1} \chi^{(m')}_0(\bbox{q}+\kappa\bbox{a}^{\ast},\omega) {\cal
  J}_{\kappa,m,m'}(\bbox{q})+1\right]^{-1}\chi^{(m)}_0(\bbox{q},\omega).\label{self1}
\end{equation}
Here $\chi^{(m)}_0(\bbox{q}, \omega)$ and
$\chi^{(m)}(\bbox{q},\omega)$ are the bare (non-interacting) and
RPA-corrected wave vector dependent dynamic  susceptibilities of
spin chains in the $m$-th sublattice, respectively. For any wave
vector $\bbox{q}$ one obtains $KM$ equations: susceptibilities at
$K$ wave vectors become coupled to each other and there are $M$
independent chain-sublattices.

For a Heisenberg AF chain, at energy transfers much smaller than
the in-chain exchange constant, $\chi^{(m)}_0(\bbox{q})$ is
typically very small except when momentum transfer along the chain
axis is close  to $q_0$, defined by $q_0a = \pi$. To a good
approximation we can thus replace Eq.~\ref{self1} with:
\begin{eqnarray}
 \chi^{(m)}(\bbox{q} \bbox{a}\sim \pi, \omega) & \approx &
  \left[\sum_{m'=1}^{M}\chi^{(m')}_0(\bbox{q},\omega)
  J_{m,m'}(\bbox{q})+1\right]^{-1}\chi^{(m)}_0(\bbox{q},\omega),\label{self2}\\
\chi^{(m)}(\bbox{qa} \neq \pi,\omega) & \approx & 0,\nonumber
\end{eqnarray}
where
\begin{equation}
J_{m,m'}(\bbox{q})\equiv \frac{1}{K}\sum_{n} \sum_{l}
\sum_{k=1}^{K}\sum_{k'=1}^K
 J(\bbox{R}_{n},m,m',l,k,k') \exp \left[ -{\rm i}\bbox{q}(\bbox{R}_n+\bbox{a}l+{\bbox a}(k'-k)/K+(\bbox{\rho}_{m'}-\bbox{\rho}_{m})) \right]
\end{equation}
is simply the Fourier transform of exchange interactions between
sublattices $m$ and $m'$. Note that there are only $M$ equations
in the system~\ref{self2}.

\section*{Appendix II: Equivalent Bravais lattice} In this section we shall derive a
useful relation between magnetic dynamic structure factors of two
spin lattices described by the same Hamiltonian, but featuring
different 3D arrangements of magnetic sites.  Lattice I is assumed
to be a simple Bravais lattice of magnetic sites and consists of
$N$ unit cells of volume $v$. The origin of the $n$-th unit cell
$\bbox{R}^{(I)}_n$ coincides with the position of the $n$-th spin:
$\bbox{r}^{(I)}_n\equiv \bbox{R}^{(I)}_n$. Unlike lattice I,
lattice II is non-Bravais, having $M$ spins per unit cell, and a
unit cell volume $V$=$Mv$. There are $K=N/M$ unit cells with
origins at $\bbox{R}^{(II)}_k$, respectively. Lattice II is
obtained from lattice I by shifting each spin $n$ by
$\bbox{\rho}_n$. In lattice II the spins are thus positioned at
$\bbox{r}^{(II)}_n\equiv \bbox{r}^{(I)}_{n} + \bbox{\rho}_n$. The
relation between the two lattices allows us to write
$\bbox{r}^{(II)}_n=\bbox{R}^{(II)}_{k}
+\bbox{r}^{(II)}_{m}=\bbox{R}^{(II)}_{k}
+\bbox{r}^{(I)}_{m}+\bbox{\rho}_m$ ( $n=kM+m$, $0\le m <M$). We
assume the spin systems to be equivalent in the sense that the
spin Hamiltonian of system II written in terms of site-spin
operators $\hat{\bbox{s}}_{n}$ is identical to that of system I.
The notation introduced above is independent of that used in the
rest of the paper or in Appendix I, and is illustrated, for the
particular case of \pb, in Fig.~\ref{transformfig}.

The Fourier transform of the total spin operator for lattice I is
defined as:
\begin{equation}
s^{(I)}(\bbox{q})=\sum_{n=0}^{N-1}s_{n}\exp[-{\rm
i}\bbox{q}\bbox{R}^{(I)}_n]=\sum_{k=0}^{K-1}\sum_{m=0}^{M-1}s_{kM+m}\exp[-{\rm
i}\bbox{q}(\bbox{R}^{(I)}_{kM}+\bbox{r}_m^{(I)})].
\end{equation}
Again, for simplicity, we have omitted the spin projection
indexes. The reverse relation is given by:
\begin{equation}
s_{kM+m}=\frac{v}{(2\pi)^3}\int d\bbox{q}
s^{(I)}(\bbox{q})\exp[{\rm i}
\bbox{q}(\bbox{R}^{(I)}_{kM}+\bbox{r}_m^{(I)})],
\end{equation}
where the integral is taken over the Brillouin zone of the Bravais
lattice ( a reciprocal-space volume $K$ times as large as the
Brillouin zone of the lattice II). Similarly, for lattice II, by
definition:
\begin{equation}
s^{(II)}(\bbox{q})=\sum_{k=0}^{K-1}\sum_{m=0}^{M-1}s_{kM+m}\exp[-{\rm
i}\bbox{q}(\bbox{R}^{(II)}_{k}+\bbox{r}_m^{(II)})].
\end{equation}
The equivalence of the two systems allows us to directly combine
the last two equations:
\begin{equation}
s^{(II)}(\bbox{q})=\frac{v}{(2\pi)^3}\sum_{k=0}^{K-1}\sum_{m=0}^{M-1}\exp[-{\rm
i}\bbox{q}(\bbox{R}^{(II)}_{k}+\bbox{r}_m^{(II)})]\int d\bbox{p}
\exp[{\rm
i}\bbox{p}(\bbox{R}^{(I)}_{kM}+\bbox{r}_m^{(I)})]s^{(I)}(\bbox{p}).\label{a1}
\end{equation}
The expression can be simplified by noting that
\begin{equation}
\sum_{k=0}^{K-1}\exp[-{\rm
i}(\bbox{q}-\bbox{p})\bbox{R}^{(II)}_{k}]=
\frac{(2\pi)^3}{V}\sum_{\bbox{T}}\delta(\bbox{p}-\bbox{q}-\bbox{T}).
\end{equation}
Here the inner sum is taken over reciprocal-lattice points
$\bbox{T}_k$, $k=0...K$, for lattice II. Plugging this result into
the expression into Eq.~\ref{a1} gives:
\begin{eqnarray}
s^{(II)}(\bbox{q}) & = & \frac{v}{V}\int d\bbox{p} \
s^{(I)}(\bbox{p})\sum_{\bbox{T}}\delta(\bbox{p}-\bbox{q}-\bbox{T})\sum_{m=1}^{M-1}\exp[-{\rm
i}(\bbox{q}\bbox{r}_m^{(II)}-\bbox{p}\bbox{r}_m^{(I)})]\nonumber\\
& = &\frac{1}{M}\sum_{m=0}^{M-1}\exp(-{\rm
i}\bbox{q}\bbox{\rho}_{m})\sum_{\mu=1}^{M-1}s^{(I)}(\bbox{q}+\bbox{T}_\mu)\exp({\rm
i}\bbox{T}_\mu\bbox{r}_m^{(I)})\label{a2}
\end{eqnarray}
In this the last expression the first sum is taken over all
reciprocal-lattice vectors $\bbox{T}_{\mu}$ of lattice II within
the first Brillouin zone of lattice I.

Equation \ref{a2} for the Fourier transform of spin operators can
now be directly plugged into the definition of dynamic structure
factors for the two lattices:
\begin{eqnarray}
S^{(I)}(\bbox{q},\omega)\equiv \frac{1}{2 \pi \hbar}\int d t
\exp(-{\rm i}\omega t) \langle
s^{(I)}(-\bbox{q},0)s^{(I)}(\bbox{q},t))\rangle\\
S^{(II)}(\bbox{q},\omega)\equiv \frac{1}{2 \pi \hbar}\int d t
\exp(-{\rm i}\omega t) \langle
s^{(II)}(-\bbox{q},0)s^{(II)}(\bbox{q},t))\rangle.
\end{eqnarray}
A straightforward, though somewhat tedious calculation gives the
result that we were after:
\begin{equation}
S^{(II)}(\bbox{q},\omega)
=\sum_{\mu=0}^{M-1}\left|\sum_{m=0}^{M-1}\exp(-{\rm
i}\bbox{q}\bbox{\rho}_m)\exp({\rm i}\bbox{T}_\mu
\bbox{r}_m^{(I)})\right|^2S^{(I)}(\bbox{q}+\bbox{T}_\mu,\omega)\label{transform}
\end{equation}


\begin{figure}
\caption{$D-J_1$ Phase diagram for weakly coupled Haldane spin
chains,\protect\cite{Sakai90} showing the location of some
well-characterized quasi-1D $S=1$ systems. For each material $J_1$
is scaled by the coordination number. CsNiCl$_3$ data from
Ref.~\protect\onlinecite{Morra88}; Y$_2$BaNiO$_5$ data from
Ref.~\protect\onlinecite{Xu96}; NENP data from
Ref.~\protect\onlinecite{Regnault94}; AgVP$_2$S$_6$ data from
Ref.~\protect\onlinecite{Mutka91}. \pb\ and \sr\ are placed based
on the results of this work.} \label{phase}\label{fig1}
\end{figure}

\begin{figure}
\caption{(a) Crystal structure on \pb\ and \sr. The magnetic
chains are formed by edge-sharing NiO$_6$ octahedra (dark gray)
and are bridged by VO$_4$ tetrahedra (light gray). (b) A
perspective view of the spiral-shaped Ni$^{2+}$-chains in \pb\ and
\sr.} \label{struct}\label{fig2}
\end{figure}

\begin{figure}
\caption{Geometry of inter-chain interactions in \pb\ and \sr\
(dark bonds). The strongest inter-chain coupling is assumed to be
between nearest-neighbor Ni$^{2+}$ ions from adjacent chains.
These sites are offset relative to each other by $c/4$ along the
chain axis.} \label{coupling}\label{fig3}
\end{figure}

\begin{figure}
\caption{Topology of magnetic interactions \pb\ and \sr.
Inter-chain bonds are shown as dashed lines. Solid lines are
in-chain Ni-Ni bonds. (a): Model used in
Ref.~\protect\onlinecite{Uchiyama99}. (b): Model used in the
present paper.} \label{topo}\label{fig4}
\end{figure}

\begin{figure}
\caption{False-color plot of inelastic neutron scattering
intensities measured in \pb\ (a) and \sr\ (b) at $T=1.5$~K using
an area-sensitive detector setup. The background has been
subtracted, as described in the text. Shaded areas indicate the
location of possible spurions.} \label{ASDdata}\label{fig5}
\end{figure}

\begin{figure}
\caption{Constant-energy scans measured in \pb\ (a) and \sr\ (b)
powder samples at $T=2$~K in the standard 3-axis mode. The solid
lines are simulations based on parameters obtained in a global fit
to the ASD data, as described in the text. Dashed lines are
similar simulations for non-interacting spiral-shaped spin
chains.} \label{conste}\label{fig6}
\end{figure}

\begin{figure}
\caption{Typical constant-$Q$ scans measured in \pb\ powder sample
at $T=2$~K in the standard 3-axis mode. The solid lines are as in
Fig.~\protect{\ref{conste}}. } \label{constqpb}\label{fig7}
\end{figure}

\begin{figure}
\caption{Typical constant-$Q$ scans measured in \sr\ powder sample
at $T=2$~K in the standard 3-axis mode. The solid lines are as in
Fig.~\protect{\ref{conste}}. } \label{constqsr}\label{fig8}
\end{figure}

\begin{figure}
\caption{Simulated inelastic scattering cross section based on
parameters obtained in a global fit to the ASD data for \pb\ (a)
and \sr\ (b).} \label{simpb}\label{fig9}
\end{figure}

\begin{figure}
\caption{Simulated inelastic scattering cross section based on
parameters obtained in a global fit to the ASD data for \pb,
assuming a reversed sign of $J_1$  (a) and no inter-chain
interactions (b).} \label{simpbbad}\label{fig10}
\end{figure}

\begin{figure}
\caption{Constant-$Q$ scans extracted from the ASD data sets for
\pb\ (a) and \sr\ (b) at a momentum transfer that corresponds to
the 3D antiferromagnetic zone-center (symbols). Unlike in \sr, the
measured inelastic intensity in \pb\ extrapolates to zero at a
non-zero energy transfer (arrows). This is even better seen in
constant-$Q$ scans measured in the standard 3-axis mode (c). Open
and solid symbols correspond to \sr\ and \pb, respectively. In the
latter plot the data for the two compounds were brought to a
single scale using the elastic incoherent signal. In all cases the
solid lines are as in Fig.~\protect{\ref{conste}}. }\label{fig11}
\label{gapnogap}
\end{figure}

\begin{figure}
\caption{Typical constant-$Q$ scans measured in \sr\ powder at
different temperatures, at a momentum transfer that corresponds to
the 3D antiferromagnetic zone-center. The solid lines are fits to
the model cross section, as described in the text. }\label{fig12}
\label{vstdata}
\end{figure}

\begin{figure}
\caption{Measured temperature dependence of the mean intrinsic
Haldane gap $\langle \Delta \rangle$ (a) and intensity prefactor
in the model cross section described in the text (b) for \sr.}
\label{vstresult}\label{fig13}
\end{figure}

\begin{figure}
\caption{Measured temperature dependence of the 3D gap energies in
for \sr. The solid lines are as in Fig.~\protect\ref{vstresult}.}
\label{vstresult2}\label{fig14}
\end{figure}

\begin{figure}
\caption{Relation between the actual structure in \pb\ (b) and the
equivalent Bravais lattice of spins (a). The notation is as in
Appendix II.} \label{transformfig}\label{fig15}
\end{figure}

\end{document}